\documentclass[preprint,superscriptaddress,preprintnumbers,amsmath,amssymb,pra]{revtex4}
\usepackage{graphicx}
\usepackage{amsmath}

\begin{document}

\thispagestyle{empty}

\title{Examining the Casimir puzzle with upgraded technique and
advanced surface cleaning}

\author{Mingyue Liu}
\affiliation{Department of Physics and Astronomy, University of California, Riverside, California 92521, USA}

\author{Jun Xu}
\affiliation{Department of Physics and Astronomy, University of California, Riverside, California 92521, USA}

\author{
G.~L.~Klimchitskaya}
\affiliation{Central Astronomical Observatory at Pulkovo of the
Russian Academy of Sciences, Saint Petersburg,
196140, Russia}
\affiliation{Institute of Physics, Nanotechnology and
Telecommunications, Peter the Great Saint Petersburg
Polytechnic University, Saint Petersburg, 195251, Russia}

\author{
V.~M.~Mostepanenko}
\affiliation{Central Astronomical Observatory at Pulkovo of the
Russian Academy of Sciences, Saint Petersburg,
196140, Russia}
\affiliation{Institute of Physics, Nanotechnology and
Telecommunications, Peter the Great Saint Petersburg
Polytechnic University, Saint Petersburg, 195251, Russia}
\affiliation{Kazan Federal University, Kazan, 420008, Russia}

\author{
 U.~Mohideen\footnote{Umar.Mohideen@ucr.edu}}
\affiliation{Department of Physics and Astronomy, University of California, Riverside, California 92521, USA}

\begin{abstract}
We performed measurements of the gradient of the Casimir force between Au-coated
surfaces of a sphere and a plate by means of significantly upgraded dynamic
atomic force microscope (AFM) based technique. By introducing combined cleaning
procedure of interior surfaces of the vacuum chamber and the test bodies by means
of UV light and Ar ions, we reached higher vacuum and eliminated the role of
electrostatic patches. Furthermore, the use of much softer cantilever allowed a
sixfold decrease of the systematic error in measuring the force gradient. The
experimental data are compared with theoretical predictions of the Lifshitz
theory taking into account corrections due to the inaccuracy of the proximity force
approximation and that due to surface roughness. It is shown that the theoretical
approach accounting for the relaxation properties of free electrons is excluded
by the data up to a larger than previous sphere-plate separation of 820~nm,
whereas an alternative approach is found in a very good agreement with the data.
Importance of these results in connection with the foundations of quantum statistical
physics is discussed.
\end{abstract}

\maketitle

 The Casimir force originally predicted \cite{1} between two uncharged ideal metal planes
was later generalized for any two material bodies and explained as an effect of the
zero-point and thermal fluctuations of the electromagnetic field along with the
van der Waals force \cite{2} (see the monograph \cite{3} for further developments).
Presently the Casimir force has been measured in numerous experiments to gain
a fundamental understanding of the physics (see the reviews \cite{4,5,6}),
 and is proposed for use
in the next generation of nanotechnological devices \cite{7,8,9,10,11,12,13}.

A comparison between the measurement data of the most precise experiments and
theoretical predictions of the Lifshits theory revealed a puzzling
inconsistency of  fundamental significance.
It was realized \cite{3,4,6} that the theory is in conflict with the data 
at separations below $1~\mu$m 
if the real part of the conductivity of materials is taken into account in
computations (at the moment there are no direct measurements of the Casimir 
interaction between metallic surfaces at separations above $1\mu$m). 
A good agreement between the theory and the data is regained if one
neglects the low-frequency relaxation properties of free charge carriers and the dc
conductivity in dielectrics (the most recent experiments demonstrating this result
are described in Refs.~\cite{14,15,16,17,18,19,20,21}).  In all these experiments
with the exception of Ref.~\cite{20} the difference between
the two alternative theoretical
predictions does not exceed 5\%. As a result, there were attempts to explain the
Casimir puzzle as the role of some unaccounted background effect, i.e., by an
additional force due to surface patches \cite{22,23,24}. Experiments with magnetic
test bodies \cite{17,18,19} (and especially Ref.~\cite{20} where the alternative
theoretical predictions diverged by up to a factor of 1000) have shown, however,
that the Casimir puzzle is not governed by the role of surface patches
and other background effects.

In this paper, we describe measurements of the gradient of the Casimir force
between Au-coated surfaces of
a sphere and a plate by means of a significantly upgraded dynamic AFM
 based technique with {\it in situ} UV and Ar ion cleaned surfaces of the test bodies.
The components of the setup reported previously \cite{16} are upgraded by
incorporating a UV lamp and an Ar-ion gun. This makes possible the removal
of contaminants from all interior surfaces
of the vacuum chamber and a significant decrease in the residual potential difference between
the test bodies (a factor of 10 compared to Ref.~\cite{16}).
Furthermore, the force sensitivity was improved by the preparation and use of
 a cantilever with a smaller (factor of 10) spring constant than in the previous work \cite{16}.
As a result, the calibration  constant in the present measurements is by an order of
magnitude larger than in Ref.~\cite{16}  leading to smaller (by a factor of 6)
systematic error in
measuring the gradient of the Casimir force. After a comparison between the measurement data
and theory, these improvements
allowed clear discrimination between two theoretical approaches mentioned
above up to the separation distance of 820~nm (compared with 420~nm in Ref.~\cite{16}).

\begin{figure}[t]
\vspace*{-10cm}
\centerline{\hspace*{4.5cm}
\includegraphics{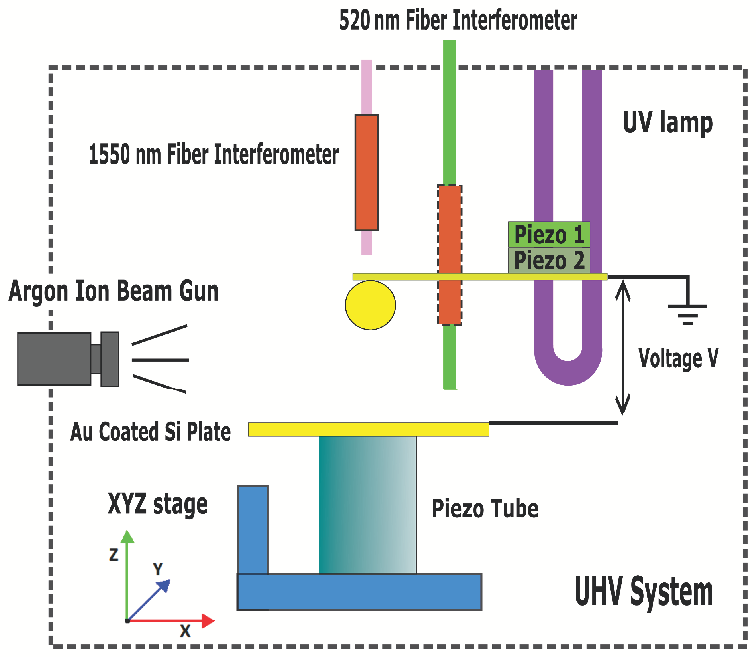}
}
\vspace*{-10.5cm}
\caption{\label{fg1}
Schematic of the upgraded experimental setup (see text for
further discussion).
}
\end{figure}

The schematic of the upgraded experimental setup is presented in Fig.~\ref{fg1}.
Below we remark only on the novel elements. Additional details of the setup and
description of the calibration procedures can be found in Refs.~\cite{16,21}.
The gradient of the Casimir force was measured between the Au-coated hollow glass
sphere of $R=43.446\pm 0.042~\mu$m radius (measured by means of scanning
electron microscope after the
experiment was completed) and Au-coated silicon plate. 
The hollow glass spheres were made from liquid phase and therefore are 
almost spherical with the difference along two perpendicular axes being less than or 
equal to 0.1\%, i.e., of the order of the radius measurement error. 
The sphere is attached to the
electrically grounded cantilever. The spring constant $k$ of this cantilever was
reduced by decreasing its thickness and width by means of etching with 60\% KOH solution
at a temperature of $50\,{}^{\circ}$C for 55 seconds. Relatively high concentration and
temperature were used to obtain smooth surfaces after etching \cite{25}.

The use of the polished silicon wafer as the base plate instead of sapphire or fused
silica plates used in previous measurements \cite{16,21}  and an $E$-beam evaporator
for making the Au coatings  instead of a thermal evaporator allowed a decrease in the
surface roughness by up to a factor of 2. The r.m.s. roughness on the sphere and  the
plate, measured after the experiment was completed, is $\delta_s=1.13~$nm and
$\delta_p=1.08~$nm (compared with $\delta_s=2.0~$nm and
$\delta_p=1.8~$nm in Ref.~\cite{16}). The Au-coated plate is mounted on a piezoelectric
tube which helps to precisely control its position (see Fig.~\ref{fg1}). In its turn,
the tube is mounted on a XYZ linear translational stage
used for the coarse approach of the plate to the sphere. The cantilever motion is
monitored with a laser optical interferometer. The laser light source has a wavelength
of 1550~nm. The second interferometer using a wavelength of 520~nm serves to measure
the movement of the Au-coated plate (see Fig.~\ref{fg1}).

The major improvement, as compared to previous experiments, is the UV followed by
Ar-ion cleaning of the test bodies and surfaces inside the vacuum chamber.
The removal of contamination on the Au sphere-plate surfaces using only Ar-ion gun
has been already studied in Ref.~\cite{21}. As a result, the residual potential
difference $V_0$ between a sphere and a plate was lowered by an order of magnitude
leading to a reduced role of electrostatic forces discussed in the literature
\cite{22,23}. In the Ar-ion cleaning, however, the ions are focused on the interacting
sphere-plate surfaces and the adsorbed contaminants on the experimental chamber walls
are not completely removed.  Thus, over longer periods of time, the desorption of contaminants
from the chamber walls leads to the redeposition of contaminant gas molecules on the
Au surfaces of test bodies resulting in increase of residual electric potential
difference $V_0$.

It has been shown \cite{26,27,28,29,30,31} that the water and organic contaminants can
be removed from the chamber walls, as well as the sphere-plate surfaces,
with the use of UV light.
We used the UV lamp (UV B-100 Water Desorption System, RBD Instruments, Inc.) with
dimensions of $10.5^{\prime\prime}$ length and $1.3125^{\prime\prime}$ diameter.
It is installed on the top of the chamber  as is
shown in Fig.~\ref{fg1}.  The emitted wavelengths are a combination of 185~nm
(2~W power) and 254~nm (5~W power). The former is absorbed by oxygen and leads to
the generation of highly reactive
ozone, whereas the letter is absorbed by most hydrocarbons and ozone
leading to their ionization and desorption from surfaces.

The combined UV/Ar cleaning procedure was performed as follows. The vacuum chamber was
first pumped down to pressure of about $9\times 10^{-9}~$Torr using the scroll mechanical
pump and the turbo pump (see Ref.~\cite{16} for a description of the vacuum system
which also includes the ion pump). Then the UV lamp was turned on for 10~min. and
because of the formation of ozone, atomic oxygen and resulting oxidation of surface
hydrocarbons to more volatile species, the chamber pressure rises to $8\times 10^{-7}~$Torr.
The volatile species were pumped out by the turbo and mechanical pumps leaving Au surfaces
and chamber walls free of water and organic contaminants. A rough measurement of $V_0$
for a sphere-plate separation of $1~\mu$m shows that the UV cleaning leads to an increase
of the residual potential difference from $49.5\pm 0.5~$mV to higher values of approximately
100--200~mV (measured after the UV lamp was turned off for 60 minutes). This can be explained by the
exposed inorganic contaminants on the sample surfaces.

To remove the latter and any remaining organics,  Ar-ion-beam bombardment \cite{21,31,32}
was used after the  UV cleaning process. The Ar ion gun was installed horizontally on the left
side of the chamber as shown in Fig.~\ref{fg1}. The sphere-plate distance was increased up
to $500~\mu$m. Note that during both UV and Ar-ion cleaning the ion pump was shut off to avoid
contamination of its electrodes. To initiate the process, the Ar gas was released into the
chamber until the pressure reached $1.2\times 10^{-5}~$Torr. The turbo pump gate valve was
half closed to keep the pressure constant. The Ar ions were accelerated under 500~V electric
field. The kinetic energy of the grazing incidence Ar ions was high enough to break chemical 
bonds of the Au oxide
and organic molecules, but low enough to prevent any sputtering of the Au surface.
The cleaning was done in 5~min. steps.
After each step the turbo pump gate valve was opened allowing the pressure to reach
$5\times 10^{-9}~$Torr in less than 30~min. Next a rough
value of $V_0$ was measured. This process
was repeated till $V_0$ reached near zero.
The final measured value was $2.0\pm 0.5~$mV. To reduce mechanical noise,
the ion pump was turned on and the turbo pump and mechanical pump were
first valved and then turned off.
Thus, a combination of {\it in situ} UV and Ar-ion cleaning allows us to achieve ultra high vacuum
$5\times 10^{-9}~$Torr resulting in clean sphere-plate surfaces with
very low and time stable
$V_0$ (the drift rate of $V_0$ was measured to be less than 0.005~mV/min).

The force gradient between the sphere and plate was measured through the shift
$\Delta\omega$ of the resonant frequency of the cantilever with attached sphere
($\omega_0$) which was recorded as a function of the sphere-plate separation $a$
in 1~nm steps. The relative sphere-plate distance $z_{\rm rel}$ was controlled by
application of voltage to the piezoelectric tube supporting the plate (see Fig.~\ref{fg1}).
The distance moved by the plate was calibrated using the interferencer fringes from
the 520~nm fiber interferometer. In doing so the absolute sphere-plate separation
$a=z_0+z_{\rm rel}$, where $z_0$ is the separation at the point of closest approach.

For the force gradient measurements, 11 voltages $V_i$ ($i=1,\,\ldots,\,11$)
were sequentially applied to the plate and the cantilever frequency shift was measured
as a function of the plate separation. The plate movement was corrected for the
mechanical drift found to be --0.00575~nm/s as described in Ref.~\cite{16}.
To subtract any systematic background in the frequency shift due to the noise coming
from the plate movement, the sphere-plate separation was increased to $50~\mu$m,
where the interaction force gradients are well beyond the experimental sensitivity,
and the experiments were repeated with a voltage $V_0$ applied to the plate.
After averaging 8 frequency shift measurements the background change in frequency
due to the mechanical movement of the plate can be found to be linear at
$1.5\times 10^{-6}~$Hz/nm. This small background noise signal due to the plate
movement was subtracted from all frequency shift signals.

\begin{figure}[b]
\vspace*{-13cm}
\centerline{\hspace*{6.cm}
\includegraphics{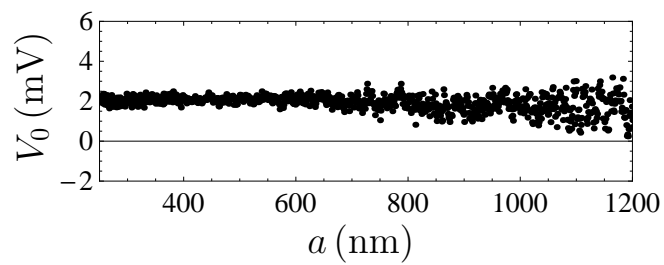}
}
\vspace*{-10.cm}
\caption{\label{fg2}
The residual potential difference between Au-coated surfaces of
 a sphere and a plate is shown by dots as a function of separation.
}
\end{figure}

The frequency shift is given by
\begin{equation}
\Delta\omega=-C\frac{\partial X(a,R)}{\partial a}(V_i-V_0)^2-
C\frac{\partial F(a)}{\partial a},
\label{eq1}
\end{equation}
\noindent
where the calibration constant $C=\omega_0/(2k)$, $X$ is the known function of the
electric force in the sphere-plate geometry \cite{3,16} and $F$ is the Casimir force.
At any given separation, $\Delta\omega$ is proportional to the square of the voltage
difference. Therefore the parameters $\gamma=C\partial{X}/\partial{a}$ and $V_0$
can be found by fitting parabolas and be plotted as functions of distance (see
Ref.~\cite{16} for details). The $V_0$ obtained at each separation are
shown in Fig.~\ref{fg2} resulting in the mean value
$\bar{V}_0=1.93\pm 0.01~$mV.
Then the best fits of the exact expression for $\gamma$ to the measured data was done
leading to $z_0=240.2\pm 0.6~$nm and $C=(6.472\pm 0.012)\times 10^5~$s/kg.
The latter is almost an order of magnitude larger than the calibration constant
$\tilde{C}=(0.683\pm 0.002)\times 10^5~$s/kg in Ref.~\cite{16}.
This increase is connected with the fact that now we use a much more sensitive cantilever with
a smaller (by a factor of 10) 
spring constant. In  previous reports performed with specially selected samples, 
possessing constant but larger $V_0$, the gradient of the total measured force was also 
larger. However, the gradients of the Casimir force obtained in Ref.~\cite{16} after a subtraction
of electrostatic contributions are in good agreement with the measurement results obtained 
here at higher precision over the wider separation region. 
  
\begin{figure}[b]
\vspace*{-3cm}
\centerline{\hspace*{-0.5cm}
\includegraphics{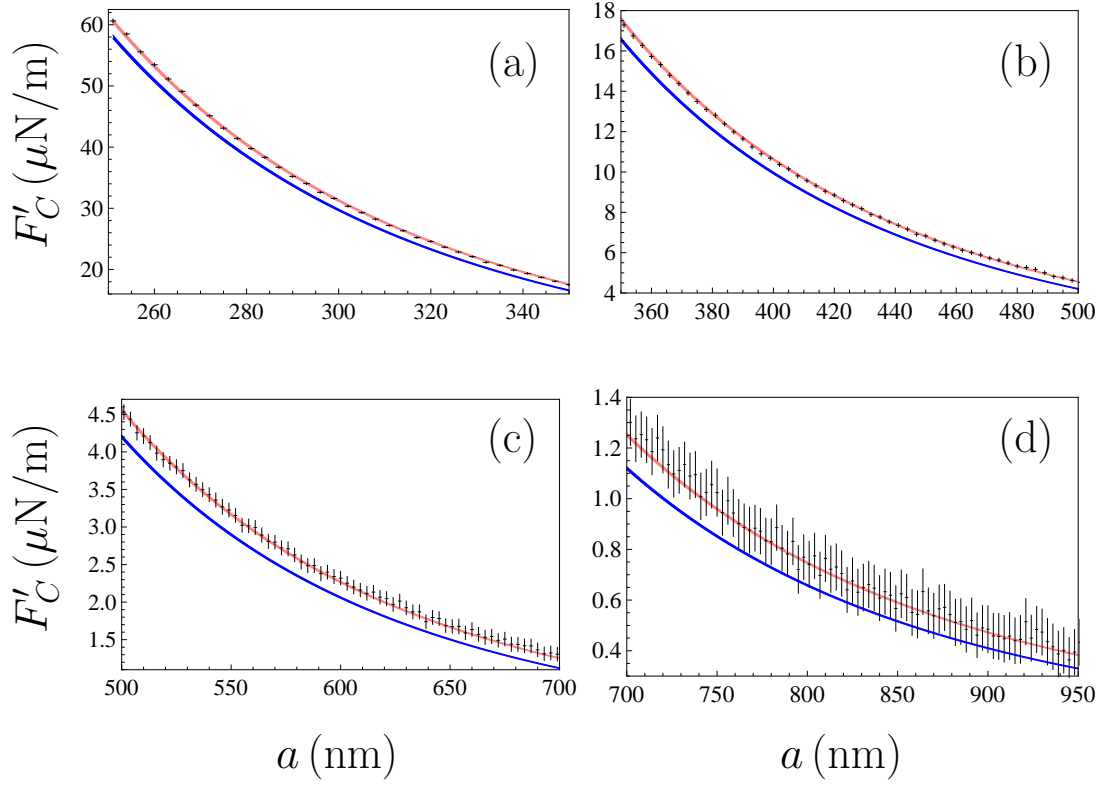}
}
\vspace*{-14.cm}
\caption{\label{fg3}
The measured gradient of the Casimir force as a function of
separation is shown as crosses. The arms of crosses indicate the
total experimental errors (for better visualization the measurement
results are shown with the step of 3~nm). Theoretical predictions
of the Lifshitz theory with neglected and included energy
losses of conduction electrons are shown as the upper and
lower bands, respectively.
}
\end{figure}

After completion of the calibration, the force gradient of the Casimir force
$F^{\prime}=\partial{F}/\partial{a}$ was calculated with a step of 1~nm using
Eq.~(\ref{eq1}) and the above parameters $C$, $z_0$ and $V_0=\bar{V}_0$.
The obtained results are presented in Fig.~\ref{fg3}(a,b,c,d) as crosses over defferent
separation regions.
The arms of the crosses show the total experimental errors determined at the 67\%
confidence level (random and systematic errors added in quadrature).
In so doing, the systematic error in the measured gradient of the Casimir force is
mostly determined by the systematic error in measuring the frequency shift which
is equal to $5.5\times 10^{-2}~$rad/s in this experiment.

The gradient of the Casimir force in the experimental configuration was computed in the
framework of the Lifshitz theory taking into account corrections to the proximity
force approximation and surface roughness. The thicknesses of the Au coatings on the sphere
and the plate were $118\pm 1~$nm and $120\pm 1~$ nm, respectively. This is more than
sufficient for these coatings to be considered as infinitely thick.
As a result, the gradient of the Casimir force is expressed as
\begin{equation}
F^{\prime}(a)=-2\pi R\left[1+\beta(a,R)\frac{a}{R}\right]
\left(1+10\frac{\delta_s^2+\delta_p^2}{a^2}\right)\,P(a),
\label{eq2}
\end{equation}
\noindent
where $P(a)$ is the Casimir pressure between two Au semispaces and the function $\beta$
quantifies corrections to the proximity force approximation determined using different
approaches in Refs.~\cite{33,33A,33B,34I,34} (here we use the computational results for $\beta$
obtained in Ref.~\cite{34I} for the force gradient).
The Casimir pressure at temperature $T$ is given by the Lifshitz formula \cite{2,3,4}:
\begin{equation}
P(a)=-\frac{k_BT}{\pi}\sum_{l=0}^{\infty}{\vphantom{\sum}}^{\prime}
\int_0^{\infty}\!\!\!\!\!q_lk_{\bot}dk_{\bot}\sum_{\alpha}
[r_{\alpha}^{-2}(i\xi_l,k_{\bot})e^{2aq_l}-1]^{-1},
\label{eq3}
\end{equation}
\noindent
where $k_B$ is the Boltzmann constant, $k_{\bot}$ is the magnitude of wave vector projection
on the plane of plates, $q_l^2=k_{\bot}^2+\xi_l^2/c^2$,
$\xi_l=2\pi k_BTl/\hbar$ ($l=0,\,1,\,2,\,\ldots$) are the Matsubara frequencies,
the prime on the summation sign divides the term with $l=0$ by 2, and the sum in $\alpha$
is over the two independent polarizations of the electromagnetic field,
transverse magnetic ($\alpha={\rm TM}$) and transverse electric ($\alpha={\rm TE}$).
The reflection coefficients are expressed via the dielectric permittivities of Au
$\varepsilon_l=\varepsilon(i\xi_l)$ calculated at the imaginary Matsubara frequencies
\begin{equation}
r_{\rm TM}(i\xi_l,k_{\bot})=\frac{\varepsilon_lq_l-k_l}{\varepsilon_lq_l+k_l},
\quad
r_{\rm TE}(i\xi_l,k_{\bot})=\frac{q_l-k_l}{q_l+k_l},
\label{eq4}
\end{equation}
\noindent
where $k_l^2=k_{\bot}^2+\varepsilon_l\xi_l^2/c^2$.

Computations with Eqs.~(\ref{eq2})--(\ref{eq4}) have been made at the $20^{\circ}$C which is
the experiment temperature. The values of $\varepsilon_l$ were obtained from the tabulated optical
data for Au extrapolated down to zero frequency either by the Drude model taking into account
the energy losses of conduction electrons or by the plasma model which neglects
these losses \cite{3,4}. The respective values for the function $\beta$ \cite{34I} have been
used in both versions of the computations. The computational results are shown in Fig.~\ref{fg3}
by the upper and lower bands computed when using the optical data extrapolated by the plasma
and Drude model, respectively. The widths of the bands reflect the theoretical errors which
are mostly determined by the inaccuracies in the optical data.

As is seen in Fig.~\ref{fg3}, the predictions of the Lifshitz theory with the inclusion
of energy losses of conduction electrons are excluded by the measurement data
up to the separation distance $a\approx 820~$nm. At the same time, the predictions of the
same theory with the energy losses of the conduction electrons neglected are in excellent
agreement with the data. The same conclusion, but at separations up to 420~nm was obtained
earlier by means of the dynamic AFM \cite{16} and at separations up to 750~nm
by means of a micromachined oscillator \cite{3,4,20,D1,D2}.

To conclude, in this experiment we reconsider the problem which remains a long
standing puzzle. In doing so, special experimental efforts have been taken to avoid
any impact of electrostatic patches on the measurement results for the Casimir force.
On the theoretical side, due to the long-term efforts of several authors it was shown that
the influence of deviations from the proximity force approximation \cite{33B,34I},
as well as the role of surface roughness \cite{3,4,35,36}, does not change the obtained results.
Thus, the Casimir puzzle assumes a fundamental importance casting doubts on the basic
assumption of quantum statistical physics that a material
system responds similarly to electromagnetic fields with nonzero field strength and 
to fluctuating  fields possessing  zero field strength and nonzero dispersion.
The complete resolution of this problem requires  measurements at large separations.
The present experiment is a step forward in this direction.

The work of M.~L., J.~X.~and U.~M.~was partially supported by the NSF
grant PHY-1607749.
M.~L., J.~X.~and U.~M.~acknowledge discussions with R.\ Schafer and Tianbai Li.
V.~M.~M.~was partially funded by the Russian Foundation for Basic
Research, Grant No. 19-02-00453 A. His work was also partically
supported by the Russian Government Program of Competitive Growth
of Kazan Federal University.

\newpage
\end{document}